\documentclass[pss]{wiley2sp} 
\usepackage{amsmath}
\begin{document}

\title{The effect of chemical disorder on the magnetic anisotropy
of strained Fe-Co films}

\titlerunning{MCA of Fe-Co Films}

\author{%
  C. Neise\textsuperscript{\textsf{\bfseries 1}},
  S. Sch\"onecker\textsuperscript{\textsf{\bfseries 1}},
  M. Richter\textsuperscript{\textsf{\bfseries 1}},
  K. Koepernik\textsuperscript{\textsf{\bfseries 1}},
  H. Eschrig\textsuperscript{\Ast,\textsf{\bfseries 1}}}

\authorrunning{C. Neise et al.}

\mail{e-mail
  \textsf{h.eschrig@ifw-dresden.de}, Phone:
  +49-351-4659-569, Fax: +49-351-4659-750}

\institute{%
  \textsuperscript{1}\,IFW Dresden, P.O.\,Box 270116, D-01171 Dresden, Germany}
\received{XXXX, revised XXXX, accepted XXXX} 
\published{XXXX} 


\abstract{%
%
%
%
\abstcol{%
Strained Fe-Co films have recently been demonstrated to
exhibit a large magnetocrystalline anisotropy (MCA) and thus
to be of potential interest as magnetic storage material.
Here, we show by means of density-functional (DF) calculations,
that chemical order can remarkably enhance the MCA.
  }{%
We also investigate the effect of relaxation perpendicular to
the applied strain and evaluate the strain energy as a function
of Co concentration and substrate lattice parameter.
On this basis, favourable preparation routes for films with a
large perpendicular anisotropy are suggested.
   }}
%
%
\maketitle   

\section{Introduction}

Materials for high-density magnetic recording media (hard disc drives)
have to obey two competing requirements~\cite{weller00}.
On the one hand, stability
against thermally activated switching is guaranteed only if the volume-integrated
MCA of a single storage bit is
larger than about $50$ $k_{\rm B}T$,
where $k_{\rm B}$ is the Boltzmann constant and $T$ is the operating
temperature~\cite{charap97}.
On the other hand, the magnetic field needed to write a bit is, apart
from demagnetisation effects,
proportional to $ K_u/ M_{\rm s}$~\cite{kronmueller91},
where $K_u$ denotes the (uniaxial) MCA energy and $M_{\rm s}$ is
the saturation magnetisation.
Since the write field is constrained by the construction of the
write head~\cite{weller00},
a large $M_{\rm s}$ of the storage material is desirable along with a large
$K_u$, the latter demand resulting from the stability requirement.

Fe-Co alloys are well-known to have a high saturation magnetisation
at ambient conditions, but their bulk MCA is minute
due to their cubic symmetry.
It is however possible to achieve high anisotropy values
without sacrificing the advantageously large magnetisation
by designing artificial structures that break the cubic symmetry.
One way to achieve this goal would be to manufacture
Fe$|$Co superlattices. It has been demonstrated experimentally and
confirmed by DF calculations, that [(110)-Fe$|$Co]$_n$
superlattices show a large in-plane magnetic anisotropy~\cite{vasko06}.
Unfortunately, advanced recording techniques require a
magnetisation orientation perpendicular to the surface,
i.e., an out-of-plane anisotropy.
A second possibility would be the preparation of ultrathin films.
Monolayers of Fe-Co on Pt(111) show a maximum out-of-plane
anisotropy energy of about $500$ $\mu$eV per atom both in DF theory
and in low temperature experiments. However, they have Curie
temperatures close to room temperature~\cite{moulas08} and
are, thus, not suited for the discussed application.

A third, recently suggested route is to fabricate
strained, bulk-like Fe$_{1-x}$Co$_x$ films by epitaxial growth on a
suitable substrate~\cite{burkert04}.
This idea relies on the fact that alloys which are cubic
in their bulk phase can be grown as metastable tetragonal
films, if they are deposited on substrates with a fourfold
symmetric surface, e.g., the (001) surface of cubic crystals.
It was shown in a recent study that the Fe$_{0.7}$Pd$_{0.3}$ alloy
can be prepared to form
$50$ nm thick, i.e., bulk-like, epitaxial films on a number of
substrates. Thus, the centred-tetragonal structure
of these films with
lattice-parameter ratios $c/a$ between $1.09$ and $1.39$
spans almost the whole range from BCC ($c/a = 1$) to FCC
($c/a = \sqrt{2}$)~\cite{buschbeck09}.

Turning back to Fe$_{1-x}$Co$_x$ alloys, the
out-of-plane values of $\Delta E_{\rm MCA} = K_u$ $(T = 0)$ predicted by
DF calculations reach up to $800$ $\mu$eV per atom for $x = 0.6$ and
$c/a$ between $1.20$ and $1.25$~\cite{burkert04}.
We will demonstrate below, that the elastic strain energy of
Fe$_{1-x}$Co$_x$ is small enough in the interesting region
of $x$ to allow the stabilisation of films with a decent
thickness. (Note, that quantum-size effects strongly influence
the MCA of thin films at least up to $10$ monolayers (ML)
or about $1.5$ nm thickness~\cite{zhang09}.)

Meanwhile, the mentioned prediction by Burkert {\em et al.} (abbreviated
below BNEH~\cite{burkert04}) has been confirmed by several
experiments~\cite{andersson06,luo07,warnicke07,yildiz09}.
These confirmations
should be viewed with caution, since
in all quantified cases the measured values of $K_u$
seem to be lower than the predictions for comparable parameter
values of $c/a$ and of $x$ by factors of two to four.
In detail, $K_u = 108 \, \mu$eV per atom
was found for Fe$_{0.5}$Co$_{0.5}|$Pd(001) films with $3$-$10$ ML
thickness, about half the predicted value~\cite{luo07}; for
[Fe$_{0.36}$Co$_{0.64}|$Pt(001)]$_n$ superlattices, only one
half~\cite{andersson06} or one quarter~\cite{warnicke07}
of the theoretical value was measured, depending on the
way of comparison (dedicated calculation for a certain
superlattice~\cite{andersson06} or evaluation of bulk data
from various superlattice data~\cite{warnicke07}).
One should note that in the latter case the anisotropy
was measured at room temperature, which explains a part
of the discrepancy.

Looking for a reason of the possibly systematic disagreement
between theory and experiment,
a first thought might blame the mentioned quantum oscillations.
However, data evaluation in two of the three mentioned cases considered
fits to a number of systems with varying film thickness,
and the third comparison involved a calculation for the specific
geometry. Another possible source of deviation might be the
local density approximation (LSDA) applied by BNEH. We argue,
that this is unlikely since in most known cases LSDA
results for $\Delta E_{\rm MCA}$ {\em underestimate} the
experimental $K_u$-values.

Here, we are going to advocate a third idea:
While the experiments in all cases known to us were performed
on presumably chemically disordered (though structurally
well-ordered) alloys, the calculations by BNEH employed the
so-called virtual crystal approximation (VCA) to simulate
the alloy.
In this approximation, the electron number is adjusted to its
correct value according to the alloy composition by allowing
the atomic nuclei to carry a non-integer charge.
In this way, a perfect chemical order is introduced.
In our particular case, all Fe and Co atoms are replaced by
only one kind of atoms with atomic charge $26 + x$.
This means, the VCA describes a {\em chemically
ordered structure} with the correct electron number of the
chemically disordered alloy.

As the main result of the present work, we will show
that strained Fe-Co films with chemical order can have a much larger
magnetic anisotropy than chemically disordered films.
Beyond this main point, we will contrast the strain-energy
landscape with the related magnetocrystalline anisotropy energy
in order to find promising preparation parameters for
films with a large perpendicular anisotropy.

The following sections contain computational details,
numerical results and their discussion, and the conclusions.

\section{Computational details}

\begin{figure}[h]%
\includegraphics*[width=\linewidth]{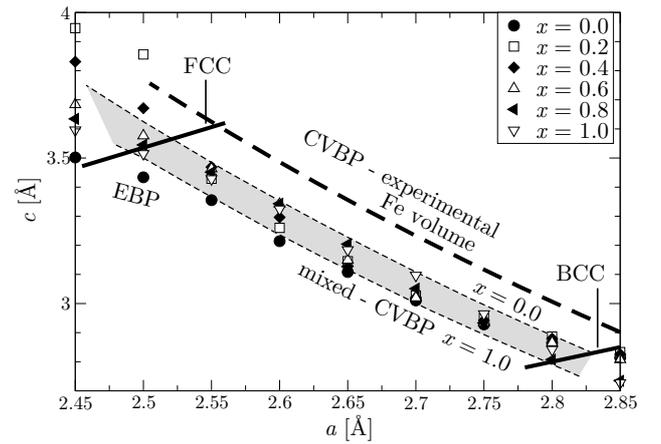}
\caption{Geometries used in the calculations: Constant volume Bain path
using the experimental volume of BCC iron (thick dashed lines, Model I) and epitaxial
Bain path calculated in VCA for different cobalt concentrations $x$ (symbols, Model II).
The shaded area with dashed margins for $x = 0$ and $x = 1$ denotes an $x$-weighted
average between the calculated volumes of BCC iron and FCC cobalt.
Thick full lines indicate $c/a$ ratios of $\sqrt{2}$ (FCC) and 1 (BCC).}
\label{fig1}
\end{figure}

The DF calculations were carried out with the all-electron,
full-potential local-orbital code FPLO, version 8.00-31~\cite{koepernik99,opahle04}.
The generalised gradient approximation (GGA) in the parameterisation by
Perdew, Burke, and Ernzerhof~\cite{pbe96} was used.
We preferred this approach against the LSDA
previously used by BNEH, since
we are interested in the evaluation of strain energies in addition
to the magnetic quantities.
The valence basis set comprised $3s$, $3p$, $3d$, $4s$, $4p$, $4d$, and $5s$ states.
Brillouin zone integrations were performed using the linear tetrahedron
method with Bl\"ochl corrections.
Structural properties and spin magnetism were evaluated
in a scalar relativistic mode, while a four-component fully relativistic
mode was used for the calculation of the MCA.

Five different structural models were employed to describe the atomistic
geometry and the chemical order
of the strained alloy films. All these models are based on the
assumptions that (i) the films are thick enough to disregard the influence
of the surface and of the substrate on the electronic structure and
that (ii) the substrate-film interaction is nonetheless strong enough
to let the in-plane lattice parameter be dictated by the substrate.
These assumptions are inherent to the concept of the epitaxial
Bain path (EBP)~\cite{bain24,alippi97} that we applied in a part of the calculations.

{\bf Model I:} A centred-tetragonal (BCT) structure with one kind of atoms
of atomic number $26 + x$ (VCA). The lattice parameter $c$ was determined
by the assumption of constant volume, where the latter was chosen to
be the experimental volume of BCC Fe at room temperature, i.e.,
$c = 2 \times 11.78$ {\AA}$^3/a^2$ (Figure 1, thick dashed line, the so-called
constant volume Bain path, CVBP).

{\bf Model II:} Same structure (BCT) and chemistry (VCA) as in Model I,
but the lattice parameter $c$ was calculated by minimisation of the
total energy $E(a,c,x)$ for $a$ and $x$ fixed, using spin polarised GGA. Related
data are given in Figure 1 with symbols for different $x$.
These data constitute the EBP. For comparison,
CVBPs are given (dashed lines including shaded area)
using the GGA volumes of BCC Fe ($x = 0$) and of FCC Co ($x = 1$).
The EBP values of $c/a$ are close to the related CVBP values except
for small $x$ in the region close to the FCC structure. This can be
understood by the strong $x$-dependence of the spin moment $\mu_s$ in this
region, see Figure \ref{fig5} below.

Contour plots presented below are based on data calculated with Model I
or II on grids with $\Delta x = 0.1$ and $\Delta a = 0.05$ {\AA}.

{\bf Model III:} L1$_0$ structure with stacks of single
quadratic Fe- and Co-layers, $x=0.5$. This structure is the
strained variant of the known $\alpha'$-Fe-Co (B2) phase, where $c/a = 1$.
The lattice parameter $c$ was calculated by
minimisation of the total energy $E(a,c)$ for $a$ fixed (EBP).

{\bf Model IV:} Stacks of single
FCC-like (Fe, Co)- and (Co, Co)-layers, $x=0.75$, and EBP condition for $c$.
This structure is derived from the L1$_2$ structure which is
obtained for $c/a=\sqrt{2}$.

{\bf Model V:} A $2\times 2\times 2$ BCT supercell with 16 atoms,
used to describe chemical disorder by means of an ensemble average.
Fe$_8$Co$_8$ ($x = 0.5$), Fe$_6$Co$_{10}$ ($x = 0.625$), and Fe$_4$Co$_{12}$
($x = 0.75$) were considered. The atoms were arranged such that the
nearest neighbour patterns match the completely disordered alloy as closely
as possible. Starting from a single atomic configuration, a symmetry-adapted
ensemble average was constructed in such a way that
its MCA correctly vanishes for $a=c$ (BCC structure). Details of the method are
described in Reference~\cite{buschbeck08}, where 32-atom supercells
modelled the case of a slightly distorted FCC structure.

The space groups $139$, $139$, $123$, $123$, $1$ were used in the calculations for
structure Models I $\ldots$ V, respectively. Structure optimisation was carried out
for Models I $\ldots$ IV, where $k$-meshes with $24$ $\times$ $24$ $\times$ $24$ points
in the full Brillouin zone (BZ) for all models were used.
For the more complex Model V,
internal relaxation is expected to be of minor importance,
since Fe and Co atoms have almost the same atomic volumes.
Thus, we abstained from a relaxation of the inner degrees of freedom.
Further, the atomic volume obtained with Model II at the appropriate
$x$-value and $a=2.65$ {\AA} (a value of particular interest, see next
section) was taken to construct a CVBP for Model V.

For the Models I $\ldots$ IV, the MCA energy was evaluated from
independent self-consistent total energy calculations for magnetic
moment orientations along the $x$-axis ($E_{100}$) and along the
$z$-axis ($E_{001}$),
\begin{equation}
\Delta E_{\rm MCA} = E_{100} - E_{001} \; .
\end{equation}
The in-plane anisotropy ($E_{100} - E_{110}$) was not considered.
In these calculations, finer $k$-meshes than in the structure
optimisations were used: $48$ $\times$ $48$ $\times$ $48$ points
in the full BZ for all models.

To cope with Model V, we relied
on the so-called magnetic force-theorem~\cite{daalderop91}:
the total energy difference $\Delta E_{\rm MCA}$
was approximated by the difference of band
energy sums, evaluated with $8$ $\times$ $8$ $\times$ $8$
$k$-points in the full BZ. In Eqn. (1) $E_{100}$ is to be replaced by
($E_{100}$ $+$ $E_{010}$)$/$$2$, where $E_{010}$ is the band energy
calculated for a magnetic moment orientation along the $y$-axis.

\section{Results and discussion}
\subsection{Strain energy}

\begin{figure}[h]%
\includegraphics*[width=\linewidth]{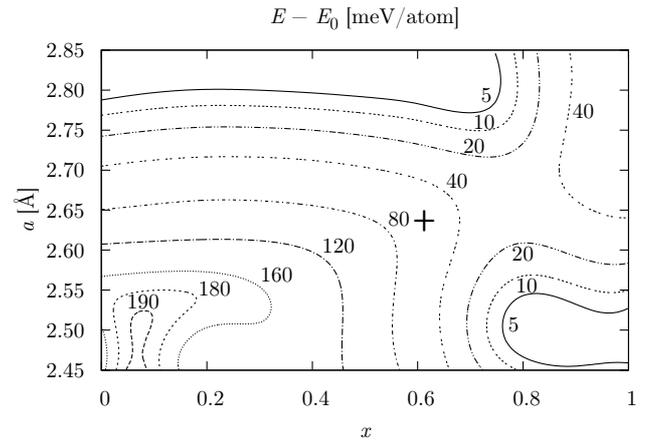}
\caption{Calculated strain energy (Model II) as a function of the
cobalt concentration $x$ and of the in-plane lattice parameter $a$.
For each value of $x$, the total energy is referred
to the lowest energy $E_0(x)$ along the respective EBP, i.e.,
there is a line $a_0(x)$ for which the presented energy is zero (not shown).
The (unstrained) ground state is BCC for $x < 0.76$ and FCC for $x > 0.76$.
(Due to interpolation from a grid the jump from BCC to FCC
at $x = 0.76$ slightly deviates from being vertical in the picture.)
The symbol + labels the point of the largest MCA
according to Fig. \protect \ref{fig2}b.}
\label{fig4}
\end{figure}

Figure \ref{fig4} shows a contour plot of the strain energy,
\begin{equation}
E_{\rm strain}(a,x) = E(a,x) - \min_a E(a,x) := E(a,x) - E_0(x) \; ,
\end{equation}
with $x$-dependent ground-state energy $E_0$,
evaluated with Model II (VCA, EBP). There are two valleys corresponding
to the BCC ($x$ $<$ $0.76$) and FCC ($x$ $>$ $0.76$) ground state,
separated by a saddle point at $x$ $=$ $0.78$; $a$ $=$ $2.66$ {\AA}.
The calculated critical Co concentration of about $76$\% for the FCC-BCC
transition is consistent with the known existence range of the
BCC-like Fe-Co B2 phase up to $72$\% Co and two-phase behaviour
(FCC + BCC) at higher Co concentration~\cite{nishizawa84}.

The strain energy
is an important quantity that determines the feasibility of
epitaxial growth. In a simplified picture, the achievable thickness
of a metastable film is inversely proportional to $E_{\rm strain}$.
Recent work on Fe$_{0.7}$Pd$_{0.3}$ films demonstrated the
possibility to grow $50$ nm films (about $300$ ML) with a strain energy of
$6$ meV per atom, estimated by DF calculations~\cite{buschbeck09}.
Considering Fe-Co films on a Rh(001) substrate ($a = 2.69$ {\AA}),
we predict a strain energy between $40$ and $80$ meV per atom for
$x < 0.65$. Indeed, experiments indicate that epitaxial growth
is stable up to about $15$ ML, while thicker films seem to become
crystallographically disordered~\cite{luo07}.

From this perspective, it seems
worthwhile to try the growth of thicker films in the saddle point region of
Figure \ref{fig4}. The related in-plane lattice parameters define a
region of very high strain (either BCC or FCC serve as zero-strain reference),
but the strain energy is comparably low.
As an additional advantage of this area the in-plane stress is low
(zero at the saddle point) which might additionally facilitate the
epitaxial growth.
We note, that in reality the saddle point area might appear at about $0.04$ {\AA}
higher $a$-values, since the GGA calculations find somewhat
smaller ground-state volumes than known from experiment,
see Figure \ref{fig1}.

\begin{figure}[h]%
\includegraphics*[width=\linewidth]{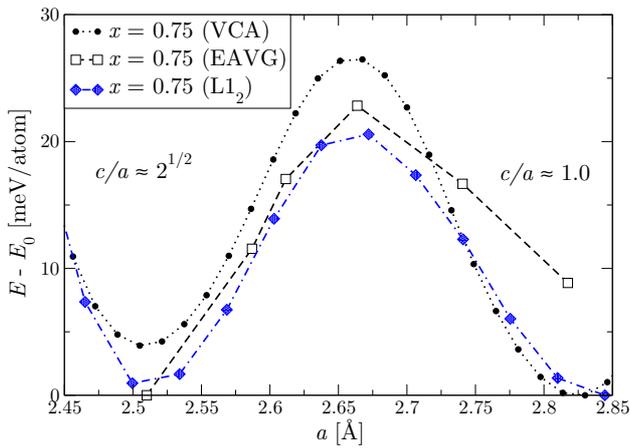}
\caption{Calculated strain energy for cobalt concentration $x = 0.75$
as a function of the in-plane lattice parameter $a$.
Results for two different
chemically ordered structures (Model II, denoted VCA, and Model IV,
denoted L1$_2$) are compared
with results for ensemble-averaged supercells simulating chemical
disorder (Model V, denoted EAVG).
The total energies are referred to the respective lowest value $E_0$($x$ $=$ $0.75$).
The VCA data are the same as those used in
Figure \ref{fig4}, but were evaluated here on a finer grid.
}
\label{fig6}
\end{figure}

To check, whether different types of structural or chemical order
have an influence on the film stability, we compare strain energies
obtained with three different structure models, Figure \ref{fig6}.
The calculations were performed for $x = 0.75$, a line cutting the
$a-x$-plane close to the saddle point.
All three models show very similar dependence of $E_{\rm strain}$
on $a$, with minima at the FCC- and BCC-(like)-structures and
a maximum (close to the saddle point in Figure \ref{fig4})
of $20$ $\ldots$ $25$ meV per atom.
We conclude, that the $a$-dependence of the
strain energy is mainly determined by the electron number
featured in the VCA.
Structural details and chemical disorder only slightly
modulate the behaviour.

\subsection{Spin moment}

\begin{figure}[h]%
\includegraphics*[width=\linewidth]{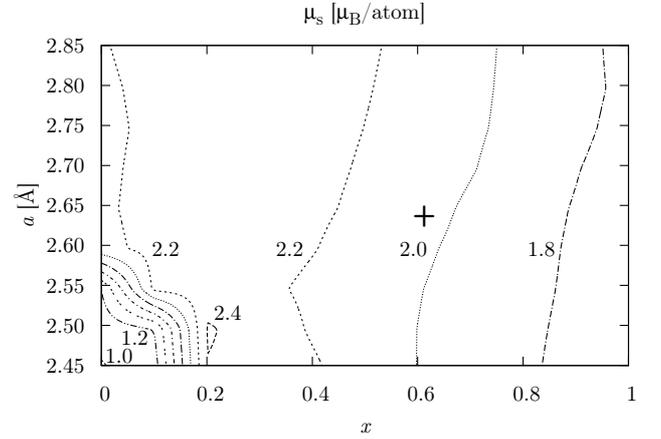}
\caption{Calculated spin magnetic moment (in VCA, Model II) as a function of cobalt
concentration $x$ and in-plane lattice parameter $a$. The symbol +
labels the point of the largest MCA according to Fig. \protect \ref{fig2}b.}
\label{fig5}
\end{figure}

For magnetic storage applications,
a large saturation magnetisation is needed to allow switching
with a magnetic field of limited strength. Figure \ref{fig5}
shows the VCA spin moment, which accounts for about $90$\%~\cite{yildiz08}
of $M_s$ $(T = 0)$, as a function of $x$ and $a$.
We find high values of $\mu_s$, depending only weakly on $x$ and being almost
insensitive to a variation of $a$, except in the region $x < 0.2; a < 2.60$ {\AA}.
The weak $x$-dependence can be understood by assuming constant atomic
moments for $x > 0.2$ and transition from weak to strong ferromagnetism
of iron in the region $0 < x < 0.2$ (Slater-Pauling behaviour).
For the case of Fe-Co films on Cu(001) ($a = 2.55$ {\AA}),
this behaviour has been confirmed in experiment~\cite{zharnikov96}.

The pronounced moment reduction in the Fe-rich area below $a = 2.60$ {\AA}
is related to the instability of ferromagnetism in FCC iron.
In the present GGA calculations, restricted to collinear ferromagnetic
states, we find a low-moment solution for very small Co concentrations
and a high-moment solution for $x \approx 0.2$.
This strong $x$-dependence of the spin moment has an effect on the
EBP (Figure \ref{fig1}) by magneto-volume coupling: for $x = 0$, the
volume is reduced in comparison to the constant-volume assumption,
while it is considerably enhanced for $x = 0.2$.

The earlier results for $\mu_s(x,c/a)$
by BNEH agree qualitatively with our data but
do not show any low-spin behaviour, since they do not include the
region very close to FCC.
Furthermore, those data were obtained for slightly larger (experimental)
volume~\cite{burkert04} which stabilises the high-moment solution.

\subsection{Magnetocrystalline anisotropy: chemically ordered films}

\begin{figure}[h]%
\includegraphics*[width=\linewidth]{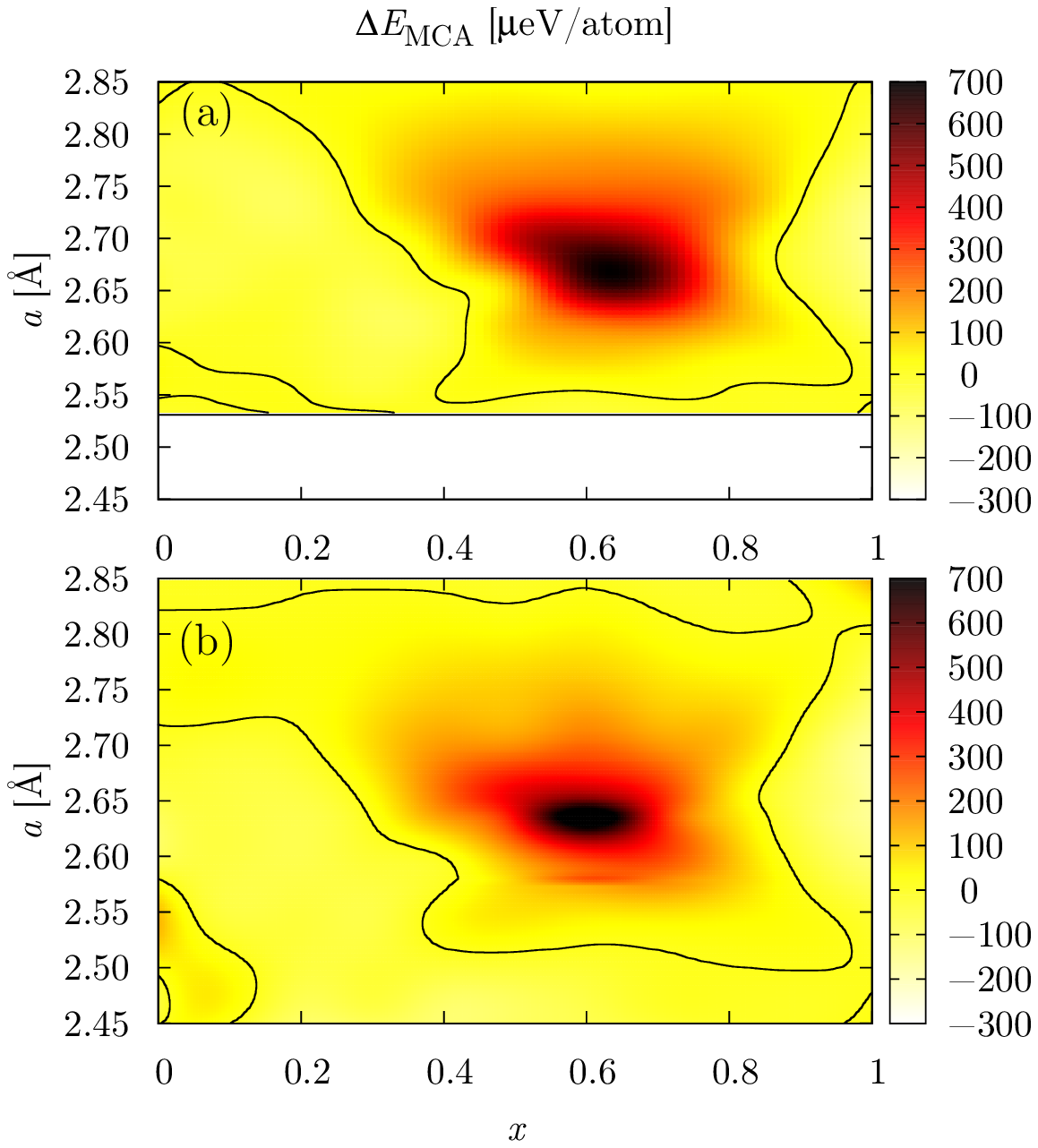}
\caption{Calculated MCA (in VCA) as a function of cobalt
concentration $x$ and in-plane lattice parameter $a$: (a) CVBP-geometry
(Model I) and (b) EBP-geometry (Model II).
Dark (brown and red) areas denote large MCA values with
an easy axis along [001], light (yellow) areas denote small MCA. Black
lines indicate zero MCA.}
\label{fig2}
\end{figure}

We now turn to the key quantity for magnetic materials applications,
the MCA. Recall, that for chemically ordered Fe-Co layers BNEH predicted
a peak in $\Delta E_{\rm MCA}$ with a maximum height of about
$800$ $\mu$eV in the vicinity of $x = 0.6$; $1.20$ $<$ $c/a$ $<$ $1.25$.
Figure \ref{fig2}a shows our data for $\Delta E_{\rm MCA} (x,a)$, evaluated
with Model I (VCA, CVBP). They agree well with the data
published by BNEH which were confirmed qualitatively by
experiments~\cite{andersson06,luo07,warnicke07,yildiz09}.

\begin{figure}[h]%
\includegraphics*[width=\linewidth]{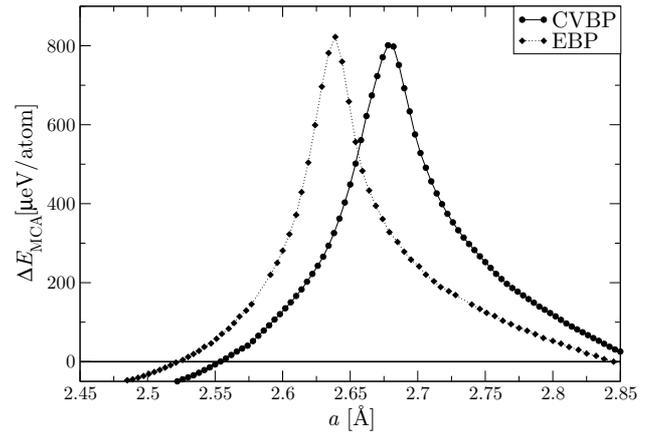}
\caption{Calculated MCA (in VCA) for cobalt concentration $x = 0.6$ as a function
of the in-plane lattice parameter $a$: comparison between CVBP
(Model I) and EBP (Model II)
(same data as in Figure \protect \ref{fig2}, but on a finer grid).}
\label{fig3}
\end{figure}

To check whether consideration of $a$- and
$x$-specific relaxation of the film perpendicular to the film plane
might affect the MCA, we repeated the calculation for the EBP (Model II,
Figure \ref{fig2}b).
By comparison of the two panels of Figure \ref{fig2} we find
that both geometries yield qualitatively the same results.
A cut at $x = 0.6$ (approximately through the peak maximum) shows,
that both structure models yield almost the same maximum MCA energy
of $800$ $\mu$eV, see Figure \ref{fig3}.
The main difference consists of a shift of the MCA-maximum from
$a = 2.68$ {\AA} (CVBP) to $a = 2.64$ {\AA} (EBP). Further, the EBP-peak has a
somewhat smaller width. In both cases, the $c/a$ ratio lies between
$1.22$ and $1.24$ (as found by BNEH). Thus, the different optimum $a$-value
originates from the volume difference (experimental vs. calculated
volume).

\subsection{Magnetocrystalline anisotropy: the effect of chemical disorder}

\begin{figure}[h]%
\includegraphics*[width=\linewidth]{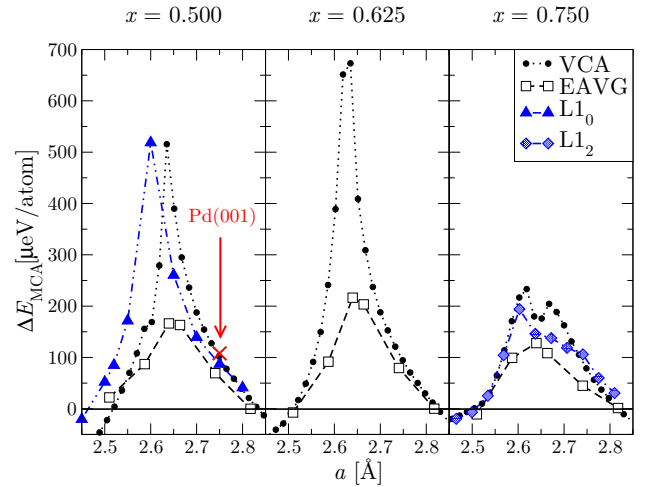}
\caption{Calculated MCA for cobalt concentrations $x = 0.5$, $0.625$,
and $0.75$ as a function
of the in-plane lattice parameter $a$: comparison between
chemically ordered structures, Models II, III, and IV
(VCA, L1$_0$, and L1$_2$-derived structure, respectively)
and ensemble-averaged supercells simulating disorder (EAVG,
Model V).
The VCA data are the same as in Figure \protect \ref{fig2}b, but were
evaluated here on a finer grid.
The red cross denotes recent experimental data
obtained for Fe$_{0.5}$Co$_{0.5}|$Pd(001)~\cite{luo07}.}
\label{fig7}
\end{figure}

Finally, we investigate how chemical disorder of the films might influence
their magnetic anisotropy. To this end, VCA results are compared with
results for two other structure types with perfect chemical order on the
one hand and with an ensemble average simulating disorder on the other
hand. We recall, that VCA is a model imposing perfect chemical order.
The calculations were performed for the most interesting
concentration range between $x = 0.5$ and $x = 0.75$.

Figure \ref{fig7} compiles calculated data for all four mentioned
structure models. Obviously, the specific type of crystallographic
structure is, in the present case, of minor importance. What matters is
chemical order. This finding is not at all trivial,
since the magnetic anisotropy is known to be sensitive to the detailed
electronic structure and, hence, to the specific crystal structure.
It was already mentioned by BNEH that L1$_0$ results for $\Delta E_{\rm MCA}$
closely resemble the VCA results for $x = 0.5$. We here confirm this point
(left panel of Figure \ref{fig7}) and add, that also by using
an L1$_2$-derived structure the VCA data ($x = 0.75$) are rather well
reproduced (right panel).

However, a striking difference is immediately visible between the
MCA of the chemically ordered structure models and the MCA of Model V,
the chemically disordered case. The latter yields $\Delta E_{\rm MCA}$
peaks smaller than those of the ordered structures by factors of $1.5$ to $3$,
depending on $x$. We suggest, that the relatively low experimental
values (compared with the prediction by BNEH) might be caused by the
lack of chemical order.

As a concluding remark, the experimental $K_u$ value for
Fe$_{0.5}$Co$_{0.5}|$Pd(001)~\cite{luo07}, red cross in the left
panel of Figure \ref{fig7}, seems to compare quite well with
our VCA or L1$_0$ data. One has to bear in mind, however, that
renormalisation of the calculated curves to the experimental volume
would shift all calculated data by $0.04$ {\AA} to the right, thus
providing coincidence of the experimental value with the simulated
disorder.

\section{Conclusions}

We have shown, that neither detailed structural relaxation
nor the specific structure type have an important influence
on the magnetic and elastic properties of strained Fe-Co films.
On the other hand, {\em chemical disorder} can reduce the
magnetic anisotropy energy by factors of $1.5$ to $3$.
We suggest that this might be a reason for the relatively
small $K_u$-values measured, hitherto.

Epitaxial growth of thick, highly strained
Fe$_{1-x}$Co$_x$ films on substrates with lattice parameter
$a$ is predicted to be most stable in the parameter range
$0.67$ $<$ $x$ $<$ $0.86$; $2.64$ {\AA} $<$ $a$ $<$ $2.76$ {\AA}, independent of chemical order.
The maximum possible values of $K_u$ are expected for
almost the same range of in-plane lattice parameters,
$2.60$ {\AA} $<$ $a$ $<$ $2.75$ {\AA} but somewhat lower Co concentration,
$0.45$ $<$ $x$ $<$ $0.70$.

Based on these findings, we suggest two promising routes
to fabricate strained Fe-Co films with large perpendicular
anisotropy:

(1) Preparation of relatively thin {\em ordered} Fe-Co L1$_0$-films
($x=0.5$) on a substrate with $a \approx 2.64$ {\AA}. The maximum thickness
of these films is probably limited to about $10$ ML by a relatively large strain
energy. On the other hand, very large $K_u$ values up to $500$ $\mu$eV
per atom are expected (Figure \ref{fig7}). Thus, a large volume-integrated
anisotropy can be achieved with relatively thin films.

(2) Preparation of relatively thick films close to the saddle point
of the strain energy, e.g., $x = 0.75$ and $2.60$ {\AA} $<$ $a$ $<$ $2.65$ {\AA}.
For these parameters, $K_u$ of chemically disordered films is limited
to about $100$ $\mu$eV per atom, too small a value to overcome the in-plane
shape anisotropy.
There is, however, a chance to enhance the magnetic anisotropy of these
films by careful annealing: calculations performed by a group including
Manfred F\"ahnle point to the possibility to stabilise
an ordered Co$_3$Fe phase by epitaxial strain~\cite{diaz06}.

We hope that our results will add another important route to the experimental
search for Fe-Co based magnetic recording media:
the preparation of {\em chemically ordered} strained films.

\begin{acknowledgement}
We thank Hongbin Zhang, Sebastian F\"ahler, and Ingo Opahle for discussion and
we thank Ingo Opahle for providing a code generating stochastic ensembles in
Model V.
\end{acknowledgement}

%
%
\bibliographystyle{pss}
\bibliography{./literature_feco}


\end{document}